\documentstyle[nato]{crckapb}

\newcommand{\half}{{\scriptstyle{{1\over 2}}}}

\def\beq{\begin{equation}}
\def\eeq{\end{equation}}
\def\bea{\begin{array}}
\def\eea{\end{array}}
\def\beqa{\begin{eqnarray}}
\def\eeqa{\end{eqnarray}}
\def\myre{{\rm Re}}

\newcommand{\re}{\relax{\rm I\kern-.18em R}}

\def\cA{{\cal{A}}}
\def\cO{{\cal{O}}}
\def\Tr{{\rm Tr}} 
\def\tr{{\rm tr}} 
\def\pl{{{\cal P}_\infty}}
\def\plo{{{\cal P}_\infty^0}}
\newcommand{\mod}{\mbox{\,mod\,}}
\newcommand{\diag}{\mbox{diag}}

\begin{opening}
\title{Instantons versus Monopoles 
\vskip-3cm\hfill {\rm INLO-PUB-21/99}\vskip1.9cm
}
\author{Pierre van Baal}
\institute{Instituut-Lorentz for Theoretical Physics\\
           University of Leiden, P.O.Box 9506\\
           NL-2300 RA Leiden, The Netherlands}
\end{opening}

\begin{document}

\begin{abstract}
We review\footnote{Presented at the workshop ``Lattice fermions
and structure of the vacuum'', 5-9 October 1999, Dubna, Russia.}{$\!$}
results of the last two years concerning caloron solutions of unit
charge with non-trivial holonomy, revealing the constituent monopole
nature of these instantons. For $SU(n)$ there are $n$ such BPS
constituents. New is the presentation of the exact values for the
Polyakov loop at the three constituent locations for the $SU(3)$
caloron with arbitrary holonomy. At these points two eigenvalues
coincide, extending earlier results for $SU(2)$ to a situation
more generic for general $SU(n)$.
\end{abstract}

\section{Introduction}
Calorons are finite temperature instanton solutions. They are
defined on $\re^3\times S^1$. Due to the periodic boundary conditions
in the time direction, the Polyakov loop at spatial infinity (the so-called 
holonomy) can take on a non-trivial value (independent of directions)
\beq
\pl=\lim_{|\vec x|\rightarrow\infty}P(\vec x),\quad
P(\vec x)={\rm P}\exp(\int_0^\beta A_0(t,\vec x)dt).
\eeq
A non-trivial value reveals that the charge one $SU(n)$ caloron actually 
contains $n$ constituent BPS monopoles, whose masses are determined by the 
eigenvalues of the Polyakov loop
\beq
\plo\equiv\exp(2\pi i\diag(\mu_1,\mu_2,\cdots,\mu_n)),\quad
\sum_{i=1}^n\mu_i=0.
\eeq
For defining the constituent masses it is important to note that one can 
choose a gauge in which $\mu_1\leq\cdots\mu_n\leq\mu_{n+1}\equiv\mu_1+1$
(for a proof see~\cite{MTP}). This guarantees that the masses $8\pi^2\nu_m
/\beta$, with $\nu_m\equiv \mu_{m+1}-\mu_m$, add up to $8\pi^2/\beta$ such 
that the action equals that of a charge one instanton.

The separation of the constituents, which are part of the moduli of 
the solution, can be chosen freely and for large separation the action 
density becomes static. This will be discussed in sect. 2, where we also
discuss the localisation of the fermion zero-mode on one of the constituents. 
In sect. 3 we give some details for the explicit computation of the gauge 
field, not presented before for $SU(n)$. New will also be the study of the 
$SU(n)$ constituent monopole location in terms of the vanishing of the 
Higgs field, which in the present context is replaced by the Polyakov loop 
variable. In the core of the constituents the Polyakov loop generically 
will have two of its eigenvalues degenerate. For $SU(2)$ this implies that 
the Polyakov loop becomes $\pm I_2$, as was verified explicitly~\cite{MTAP}. 
In sect. 4 we present for $SU(3)$ the simple result for the Polyakov loop at 
the three constituent locations. We also review the fact that in a suitable 
gauge only one of the constituents' gauge fields is non-static, in accordance 
to the Taubes-winding~\cite{Tau} required to form out of monopoles a four 
dimensional gauge field configuration with non-trivial topological charge, 
as is discussed in some detail. 

We conclude in sect. 5 with some comments concerning the fact that perhaps 
monopoles are more fundamental building blocks, since we know how to make 
instantons out of them. But we hasten to say the opposite point of view can 
be taken as well, since monopoles can be made out of an array of instantons. 
It more seems to imply they occur on equal footing, in accordance with the
``democratic'' vacuum model described in ref.~\cite{LAT97}.

\section{Densities}
Using the classical scale invariance we can always arrange $\beta=1$, as will 
be assumed throughout. A remarkably simple formula for the $SU(n)$ action 
density exists~\cite{PLBN} 
\beqa
\Tr F_{\alpha\beta}^{\,2}(x)=\partial_\alpha^2\partial_\beta^2\log\psi(x),\quad
\psi(x)=\half\tr(\cA_n\cdots \cA_1)-\cos(2\pi t),\\
\cA_m\equiv\frac{1}{r_m}\left(\!\!\!\bea{cc}r_m\!\!&|\vec\rho_{m+1}|
\\0\!\!&r_{m+1}\eea\!\!\!\right)\left(\!\!\!\bea{cc}\cosh(2\pi\nu_m r_m)\!\!&
\sinh(2\pi\nu_m r_m)\\ \sinh(2\pi\nu_m r_m)\!\!&\cosh(2\pi\nu_m r_m)\eea
\!\!\!\right),\nonumber
\eeqa
with $r_m\equiv|\vec x-\vec y_m|$ and $\vec\rho_m\equiv\vec y_m-\vec y_{m-1}$,
where $\vec y_m$ is the location of the $m^{\rm th}$ constituent monopole with
a mass $8\pi^2\nu_m$. Note that the index $m$ should be considered mod $n$, 
such that e.g. $r_{n+1}=r_1$ and $\vec y_{n+1}=\vec y_1$ (but note $\mu_{n+1}=
1+\mu_1$). It is sufficient that only one constituent location is far 
separated from the others, to show that one can neglect the $\cos(2\pi t)$ 
term in $\psi(x)$, giving rise to a static action density in this 
limit~\cite{PLBN,LAT98}.

These generalised caloron solutions can be found~\cite{KrvB,PLBN} using the 
Nahm transformation~\cite{Nahm} and the Atiyah-Drinfeld-Hitchin-Manin (ADHM) 
construction~\cite{ADHM}, related by a suitably defined Fourier transformation. 
Other methods, relying more exclusively on the Nahm transformation, were 
developed as well~\cite{Lee}.

The Nahm equation for charge one calorons reduces to an abelian problem on the 
circle, parametrised by $z\mod 1$, 
\beq
\frac{d}{dz}\hat A_j(z)=2\pi i\sum_m\rho_m^j\delta(z-\mu_m),
\eeq
giving $\hat A_j(z)=2\pi i y_m^j$, for $z\in[\mu_m,\mu_{m+1}]$. In the monopole
literature $\hat A_j(z)$ is usually denoted by $T_j(z)$. 

The basic ingredient in the construction of caloron solutions is a Green's 
function, defined on the circle $z\in[0,1]$, satisfying 
\beq
\left(\left(\frac{1}{2\pi i}\frac{d}{dz}-t\right)^2+r^2(x;z)+\frac{1}{2\pi}
\sum_m\delta(z-\mu_m)|\vec\rho_m|\right)\hat f_x(z,z')
=\delta(z\!-\!z'),
\eeq
where $r^2(x;z)\!=\!r_m^2(x)$ for $z\in[\mu_m,\mu_{m+1}]$. 
This can be solved using a similarity with the impurity scattering 
problem on the circle~\cite{PLBN,MTP}, which we present here for the 
case that $\mu_m\leq z'\leq z\leq\mu_{m+1}$ (extended to $z<z'$ by 
$\hat f_x(z',z)=\hat f_x^*(z,z')$)
\beqa
\hat f_z(z,z')=\frac{\pi e^{2\pi i t(z-z')}}{r_m\psi}\Bigl(
e^{-2\pi it}\sinh\left(2\pi(z-z')r_m\right)+\nonumber\\
<\!v_m(z')|\cA_{m\!-\!1}\cdots \cA_1\cA_n\cdots \cA_m|w_m(z)\!>\!\Bigr),
\eeqa
where the spinors $v_m$ and $w_m$ are defined by
\beqa
&v_m^1(z)=-w_m^2(z)=\sinh\left(2\pi(z\!-\!\mu_m)r_m\right),\nonumber\\
&v_m^2(z)=\hphantom{-}w_m^1(z)=\cosh\left(2\pi(z\!-\!\mu_m)r_m\right).
\eeqa

With the gauge field in the periodic gauge ($A_\alpha(t+1,\vec x)=A_\alpha(t,
\vec x)$) one can look for the fermion zero-modes that satisfy the boundary 
condition $\Psi_z(t+1,\vec x)=\exp(2\pi i z)\Psi_z(t,\vec x)$. To obtain the 
finite temperature fermion zero-mode one puts $z=\half$, whereas for 
the fermion zero-mode with periodic boundary conditions one takes $z=0$. 
For the density of these fermion zero-modes a very simple result in terms 
of the Green's function can be derived~\cite{MTCP,MTP} 
\beq
|\Psi_z(x)|^2=-(2\pi)^{-2}\partial_\alpha^2\hat f_x(z,z).
\eeq
From this it is easily seen that in case of well separated constituents the 
zero-mode is localised only at $\vec y_m$ for which $z\in[\mu_m,\mu_{m+1}]$. 
To be specific, in this limit $\hat f_x(z,z)=\pi\tanh(\pi r_m\nu_m)/r_m$
for $SU(2)$, and more generally
\beq
\hat f_x(z,z)=\frac{2\pi\sinh[2\pi(z-\mu_m)r_m]\sinh[2\pi(\mu_{m+1}
-z)r_m]}{r_m\sinh[2\pi\nu_mr_m]}.
\eeq
We illustrate the localisation of the fermion zero-modes for a typical
$SU(3)$ caloron in figure 1.

\begin{figure}[htb]
\vspace{5.5cm}
\includegraphics{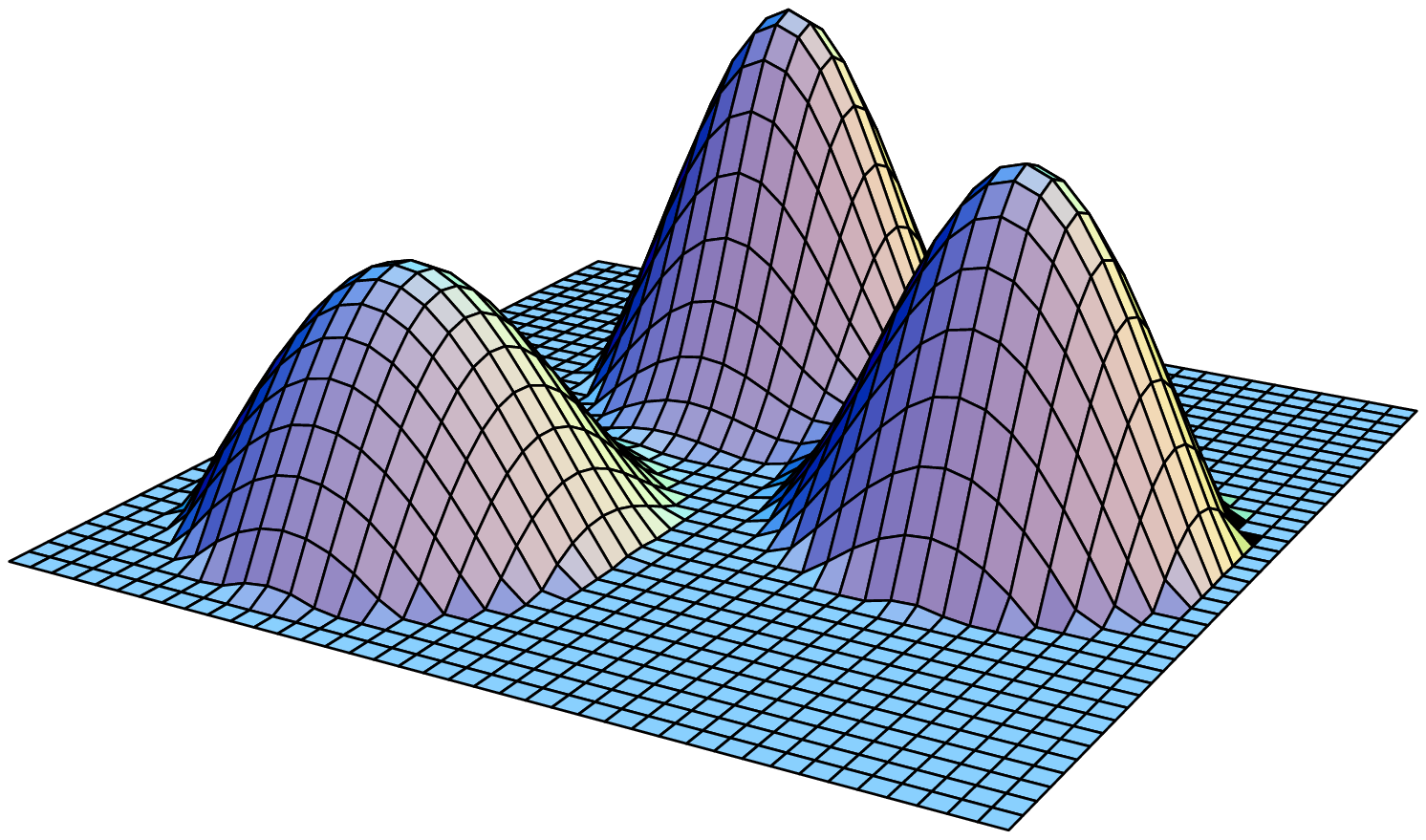}
\includegraphics{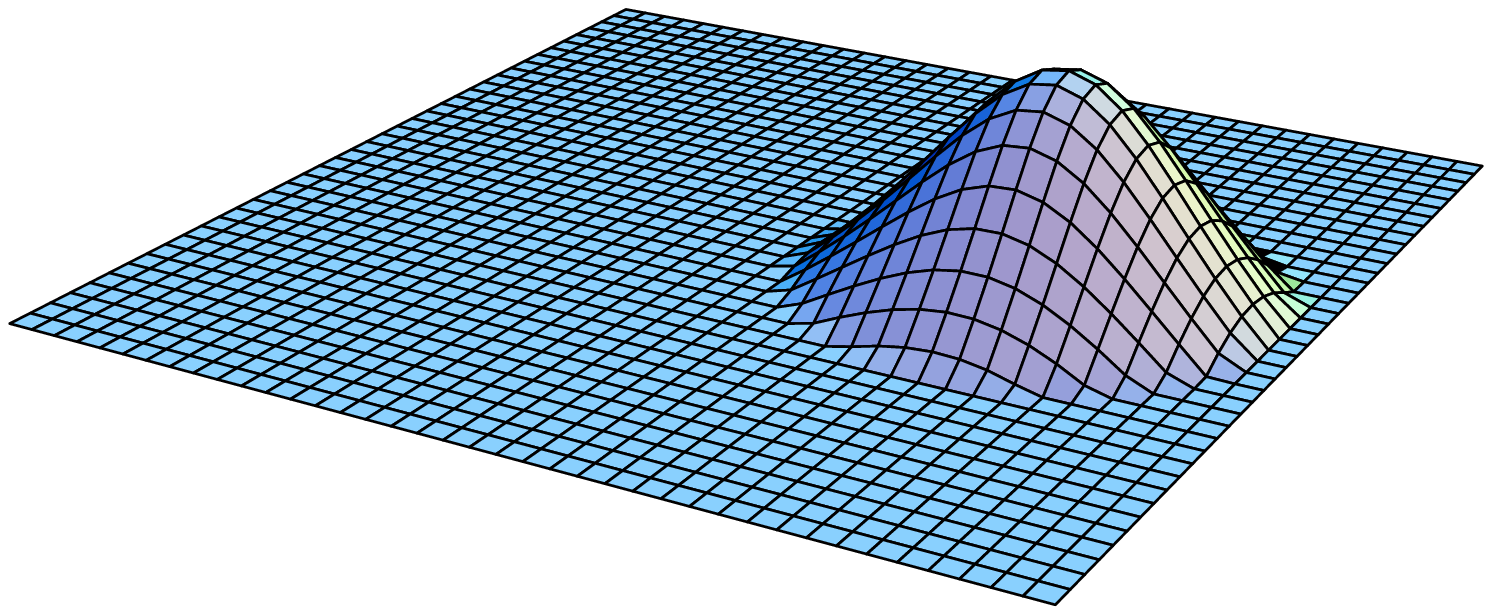}
\includegraphics{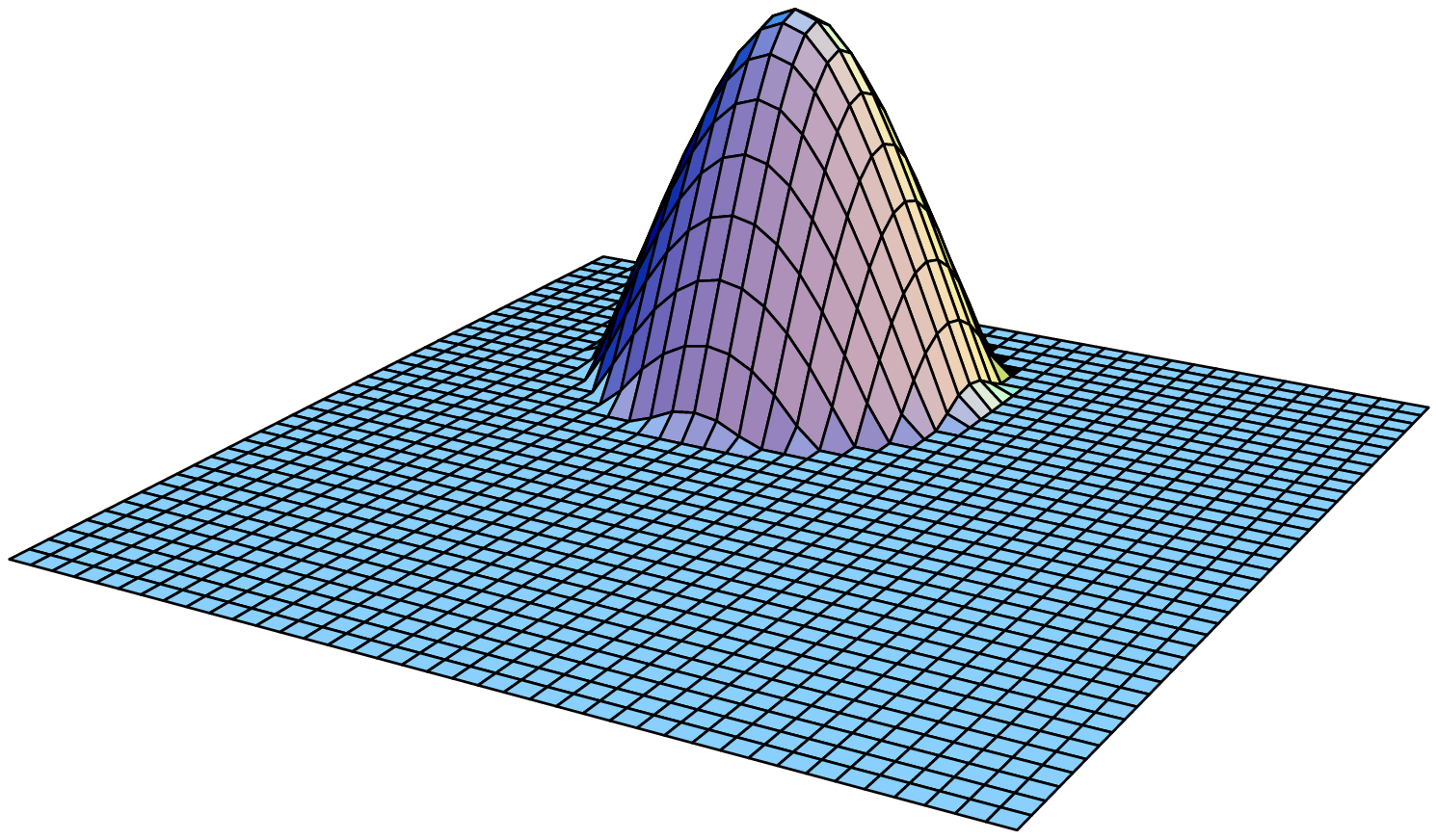}
\caption{The action density (top) for a $SU(3)$ caloron at $t=0$, for $\beta=1$,
on a logarithmic scale cut off at $1/(2e)$, with $(\mu_1,\mu_2,\mu_3)=
(-17,-2,19)/60$ shown in the plane defined by $\vec y_1=(-2,-2,0)$, $\vec y_2=
(0,2,0)$ and $\vec y_3=(2,-1,0)$. The masses $8\pi^2\nu_i$ are given by
$(\nu_1,\nu_2,\nu_3)=(0.25,0.35,0.4)$. On the bottom-left is shown
the zero-mode density for fermions with anti-periodic boundary conditions
in time and on the bottom-right for periodic boundary conditions, at equal
logarithmic scales cut off below $1/e^5$.}
\end{figure}

\section{Gauge fields}
To construct the gauge potential $A_\alpha(x)$, it is instructive to summarise 
the ADHM formalism for $SU(n)$ charge $k$ instantons~\cite{ADHM}. It employs 
a $k$ dimensional vector $\lambda=(\lambda_1,\ldots,\lambda_k)$, where 
$\lambda_i^\dagger$ is a two-component spinor in the $\bar n$ representation 
of $SU(n)$. Alternatively, $\lambda$ can be seen as an $n\times 2k$ complex 
matrix. In addition one has four complex hermitian $k\times k$ matrices 
$B_\alpha$, combined into a $2k\times2k$ complex matrix $B=\sigma_\alpha 
B_\alpha$, using the unit quaternions $\sigma_\alpha=(1_2,i\vec\tau)$ and 
$\bar\sigma_\alpha=(1_2,-i\vec\tau)$, where $\tau_i$ are the Pauli matrices. 
Together $\lambda$ and $B$ constitute a $(n+2k)\times 2k$ dimensional matrix 
$\Delta(x)$, to which is associated a complex $(n+2k)\times n$ dimensional 
normalised ($v^\dagger(x)v(x)=I_n$) zero-mode matrix $v(x)$,
\beq
\Delta^\dagger(x)v(x)\equiv\left(\lambda^\dagger,B^\dagger(x)\right)v(x)=0,
\quad\quad B(x)\equiv B-x I_k.
\eeq
Here $x$ denotes the quaternion $x=x_\alpha\sigma_\alpha$. The gauge field is 
now given by $A_\alpha(x)=v^\dagger(x)\partial_\alpha v(x)$. For this to be 
self-dual, $\Delta(x)$ has to satisfy the quadratic ADHM constraint, 
\beq
\Delta^\dagger(x)\Delta(x)=\sigma_0 f^{-1}_x,
\eeq
defining $f_x$ as a hermitian $k\times k$ Green's function. The gauge field
can conveniently be written as (cmp.~\cite{Temp,PLBN})
\beq
A_\alpha(x)=\half\phi^\half(x)\lambda\bar\eta_{\alpha\beta}\partial_\beta f_x
\lambda^\dagger\phi^\half(x)+\half[\phi^{-\half}(x),\partial_\alpha
\phi^\half(x)].
\eeq
where $\phi(x)$ is a {\em positive definite} $n\times n$ matrix, and 
$\bar\eta_{\alpha\beta}$ and $\eta_{\alpha\beta}$ are the (anti-)self-dual 
't Hooft tensors,
\beq
\phi(x)\equiv\left(1-\lambda f_x\lambda^\dagger\right)^{-1},\ 
\eta_{\alpha\beta} =\eta_{\alpha\beta}^a\sigma_a\equiv\sigma_{[\alpha}
\bar\sigma_{\beta]},\ \bar\eta_{\alpha\beta}=\bar\eta_{\alpha\beta}^a
\sigma_a\equiv\bar\sigma_{[\alpha}\sigma_{\beta]}.
\eeq

The charge one caloron with Polyakov loop $\plo$ at infinity is built out of a 
periodic array of instantons, twisted by $\plo$. This is implemented in the 
ADHM formalism by $\lambda_{p+1}=\plo\lambda_p$, which implies $\lambda^m_p=
\exp(2\pi ip\mu_m)\zeta_m$, with $\zeta$ constant, $m$ the colour and $p$ 
the ``charge'' index (spinor indices are implicit throughout), where 
$\vec\rho_m\equiv-\pi\zeta_m^{\hphantom{\dagger}}\vec\tau\zeta_m^\dagger$. 
The phases of $\zeta_m$ are related to global gauge transformations that
leave $\plo$ invariant, and define the ``framing'' of the caloron.
We note that $\hat f_x(z,z')$ is the Fourier transform of the infinite 
dimensional matrix $f_x$ as it occurs in the ADHM construction. It can be 
shown~\cite{Osb} that $\Tr F_{\alpha\beta}^{\,2}(x)=-\partial_\alpha^2
\partial_\beta^2\log\det f_x$. With the help of eq.~(6) we find 
$\partial_\alpha\log\det f_x=-\partial_\alpha\log\psi(x)$, see eq.~(3). 
We also perform Fourier transformation to obtain
\beq
\hat\lambda(z)=\sum_p e^{-2\pi piz}\lambda_p,\quad
\hat\lambda^m(z)=\delta(z-\mu_m)\zeta^m.
\eeq
This implies a remarkably simple result for the gauge field in the 
algebraic (or singular) gauge, $A_\alpha(t+1,\vec x)=\plo A_\alpha(\vec x)\ 
(\plo)^\dagger$,
\beq
A_\alpha(x)=\half\phi^\half(x)C_\alpha(x)\phi^\half(x)+\half[\phi^{-\half}(x),
\partial_\alpha\phi^\half(x)],
\eeq
with
\beq
C_\alpha^{mk}(x)\equiv\zeta_m^{\hphantom{\dagger}}\bar\eta_{\alpha\beta}
\zeta_k^\dagger\partial_\beta\hat f_x(\mu_m,\mu_k),\quad \phi^{-1}_{mk}=
\delta_{mk}-\zeta_m^{\hphantom{\dagger}}\zeta_k^\dagger\hat f_x(\mu_m,\mu_k).
\eeq

Note that $\zeta_m^\dagger\zeta_m^{\hphantom{\dagger}}=(|\vec\rho_m|-
\vec\rho_m\cdot\vec\tau)/2\pi$ and that $\sum_m\vec\rho_m=\vec 0$ implies 
a constraint on $\zeta$. In particular for $SU(2)$ one finds
$\zeta_1^{\hphantom{\dagger}}\zeta_2^\dagger=0$, and together with 
$\hat f_x(\mu_1,\mu_1)=\hat f_x(\mu_2,\mu_2)$,
$\phi(x)$ is found to be a multiple of the identity, such that 
$A_\alpha(x)=\half\phi(x)C_\alpha(x)$. The computation of $C_\alpha(x)$ 
further simplifies when rotating $\vec\rho_1=-\vec\rho_2=(0,0,\pi\rho^2)$,
which can be obtained from $\zeta_1=(1,0)\rho$ and $\zeta_2=(0,1)\rho$.
With $\chi\equiv\rho^2\hat f_x(\mu_2,\mu_1)$ and $\phi(x)=(1-\rho^2
\hat f_x(\mu_2,\mu_2))^{-1}$ this gives the result of ref.~\cite{KrvB}
\beq
A_\alpha=\frac{i}{2}\bar\eta^3_{\alpha\beta}\tau_3\partial_\beta\log\phi+
\frac{i}{2}\phi\myre\left((\bar\eta^1_{\alpha\beta}-i\bar\eta^2_{\alpha\beta})
(\tau_1+i\tau_2)\partial_\beta\chi\right).
\eeq

Also for $SU(3)$, $\hat f_x(\mu_m,\mu_n)$ can be determined from the explicit
result in eq.~(6) (since two intervals on the circle, partitioned in three
parts, always are neighbours). Choosing the three constituents to lie
in the $x$-$y$-plane we can take $\zeta_m=(|\vec\rho_m|,i\rho_m^2-\rho_m^1)/
\sqrt{2\pi|\vec\rho_m|}$. It is then a simple matter to compute the gauge 
field explicitly, albeit in a complicated form, due to the need to 
diagonalise $\phi(x)$. 

\section{Polyakov loops and Taubes-winding}
In the appendix of ref.~\cite{MTAP} the following exact result was found 
for the $SU(2)$ Polyakov loop along the line (taken along the $z$-axes)
connecting the two constituents,
\beq
\half\Tr\,P(0,0,z)=-\cos(\nu_1\pi+\half\partial_z{\rm acosh}[\half
\tr(\cA_2\cA_1)]).
\eeq
From this it is easily seen that $P(\vec x)$ takes on each of the values 
$\pm I_2$ only once, with $P(\vec y_2)=-I_2$ and $P(\vec y_1)=I_2$ for
well separated constituents. These are the equivalent to the conditions 
for the Higgs field to vanishing, an alternative way to specify the 
location of the constituents. When the constituents get nearer to each
other, these ``zeros'' shift outwards (whereas the maxima of the energy 
density shift inwards), see figure 2. 
\begin{figure}[htb]
\vspace{5.2cm}
\includegraphics{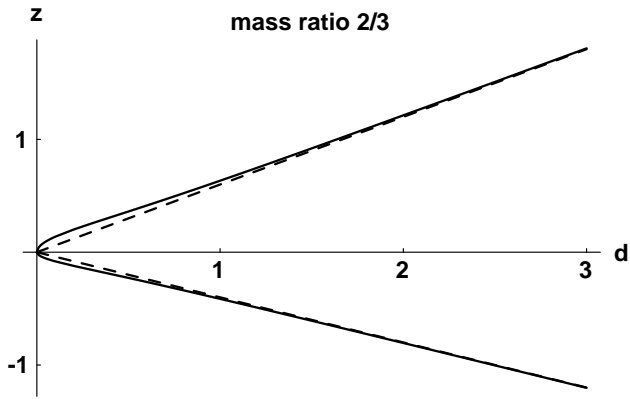}
\caption{Shift of the locations where $P^2(\vec x)=I_2$ as compared to the
location of the constituent monopole centers $\vec y_i$ for $SU(2)$. 
Horizontally is plotted the distance $d=\pi\rho^2$ between the constituents 
and vertically the position of $z_1=\nu_2d$ and $z_2=-\nu_1d$, and the 
locations where $P(0,0,z)=I_2$ ($z>0$) and $P(0,0,z)=-I_2$ ($z<0$).}
\end{figure}

Note that the result for the Polyakov loop associated to the constituent at
$\vec y_1$ is consistent with $A_0=0$, but for the constituent associated to
$\vec y_2$ this only holds after applying a gauge transformation that is
{\em anti-periodic} in the time direction (since such a gauge transformation
changes the sign of the Polyakov loop). This gauge transformation has an 
interesting relation to the so-called Taubes-winding~\cite{Tau}, which is
most easily understood by looking at the explicit expression for the gauge 
field.  In the {\em periodic gauge} one finds~\cite{KrvB}
\beqa
A_\alpha^{\rm per}(x)&=&\frac{i}{2}\bar\eta^3_{\alpha\beta}\tau_3\partial_\beta
\log\phi+\pi i\nu_1\tau_3\delta_{\alpha,0}\\&+&\frac{i}{2}\phi\myre
\left((\bar\eta^1_{\alpha\beta}-i\bar\eta^2_{\alpha\beta})
(\tau_1+i\tau_2)(\partial_\beta+2\pi i\nu_1\delta_{\beta,0})
\tilde\chi\right),\nonumber
\eeqa
where $\chi$ has the following expansion
\beq
\tilde\chi\equiv e^{-2\pi it\nu_1}\chi=\frac{4\pi\rho^2}{(r_2+r_1+\pi\rho^2)^2}
\left\{r_2e^{-2\pi r_2\nu_2}e^{-2\pi it}\!+r_1 e^{-2\pi r_1\nu_1}\right\},
\eeq
up to relative errors $\cO(e^{-4\pi\,{\rm min}(r_1\nu_1,r_2\nu_2)})$.
For large $\rho$, $\phi(x)$ becomes time independent, confirming the
static nature of the configuration for large constituent separations.
The time dependence of the constituent at $\vec y_2$ is a full (gauge) 
rotation - the {\em Taubes-winding} - responsible for the topological 
charge of the otherwise time independent monopole pair~\cite{Tau},
see figure 3. This gauge rotation is achieved by the anti-periodic 
gauge transformation $g(x)=\exp(-\pi i t\tau_3)$, since 
$g(x)(\tau_1+i\tau_2)g^\dagger(x)=\exp(-2\pi i t)(\tau_1+i\tau_2)$. Also 
note that for the spherically symmetric Bogomol'nyi-Prasad-Sommerfield (BPS) 
monopole~\cite{BPS} a gauge rotation is equivalent to an ordinary rotation. 

\begin{figure}[htb]
\vspace{2.5cm}
\includegraphics{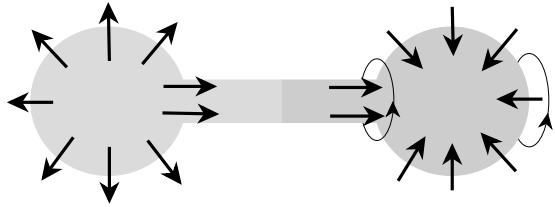}
\caption{The gauge field with unit topological charge is constructed from two 
oppositely charged monopoles by rotating one of them over one full rotation,
while moving over one time-period.}
\end{figure}

The information that the constituent at $\vec y_1$ is associated to a 
static BPS monopole, whereas the constituent at $\vec y_2$ has acquired 
a Taubes-winding can be extracted in a way~\cite{MTP} that is also valid for 
$SU(n)$. Let us first observe that taking one interval $z\in[\mu_m,\mu_{m+1}]$ 
in isolation, applying the Nahm transformation~\cite{Nahm} gives a single 
{\em static} BPS monopole located at $\vec y_m$ with mass proportional to 
the length ($\nu_m$) of the interval. Taking $|\vec y_n|\rightarrow\infty$ 
creates an infinite barrier for the interval $[\mu_n,\mu_{n+1}]$ and this 
leaves the interval $[\mu_1,\mu_n]$, allowing for the interpretation of an 
$SU(n)$ monopole with $\mu_i$ specifying the eigenvalues of the Higgs field 
at infinity, for which it is crucial the $\mu_i$ add to zero. Indeed, in 
the periodic gauge $A_0$ tends to a constant at spatial infinity. 

This {\em static} monopole solution is composed out of $n-1$ basic BPS 
monopoles of mass $\nu_m$, located at $\vec y_m$, for $m=1,\cdots,n-1$. 
We conclude that with our choice of parameters, always the field 
associated with the constituent at $\vec y_n$ has to have a time-dependence
to give rise to the topological charge of the caloron, and we conclude it is
this one that carries the Taubes-winding. Note that our argument also 
demonstrates that for $|\vec y_m|\rightarrow\infty$ with $m\neq n$, one 
is left with a gauge field that cannot be time independent, even though 
the resulting action density is static~\cite{LAT98}.

The question now arises, what is the equivalent of the zeros of the Higgs 
field for $SU(n)$. To answer this question we remember that the constituents 
are {\em basic} BPS monopoles, which are obtained by embedding $SU(2)$ in 
$SU(n)$. The $SU(2)$ subgroup relevant for this embedding is exactly 
determined by the unbroken $SU(2)$ at the core of the constituent. The 
restoration of the $SU(2)$ symmetry arises due to the degeneracy of two of 
the eigenvalues of the Higgs field, or of $P(\vec x)$ in our case (for 
$n=2$ this indeed implies a vanishing Higgs field, which has zero trace, 
or in our case it implies $P(\vec x)=\pm I_2$ as its determinant is unity). 

We denote by $\vec z_m$ the position associated to the the $m$-th constituent 
where two eigenvalues of the Polyakov loop coincide. Arranging by a global
gauge rotation that $\pl=\plo$ (see eq.~(2)), we established numerically 
for $SU(3)$ that with {\em any} choice of holonomy and constituent locations
\beqa
P_1=P(\vec z_1)=\diag(\hphantom{-}e^{-\pi i\mu_3},\hphantom{-}
                    e^{-\pi i\mu_3},\hphantom{-}e^{2\pi i\mu_3}),\nonumber\\
P_2=P(\vec z_2)=\diag(\hphantom{-}e^{2\pi i\mu_1},\hphantom{-}
                    e^{-\pi i\mu_1},\hphantom{-}e^{-\pi i\mu_1}),\\
P_3=P(\vec z_3)=\diag(-e^{-\pi i\mu_2},\hphantom{-}e^{2\pi i\mu_2},
                  -e^{-\pi i\mu_2}).\nonumber
\eeqa
For the caloron presented in figure 1 we find $|\vec z_m-\vec y_m|<0.02$,
and for well separated constituents $\vec z_m=\vec y_m$. The salient 
features are that $P_m$ is determined by $\mu_k$ {\em not} associated to 
the interval $[\mu_m,\mu_{m+1}]$ (on which $\hat A_j$ takes the value 
$2\pi iy_m^j$, see eq.~(4)). Furthermore, the extra minus signs occurring 
in $P_3$ reflect the Taubes-winding, read-off from the gauge field in the 
core of the constituent. It will not be difficult to {\em conjecture} the 
values of $P_m$ for general $SU(n)$: $P_1=\diag(e^{\pi i(\mu_1+\mu_2)},
e^{\pi i(\mu_1+\mu_2)},e^{2\pi i\mu_3},\cdots,e^{2\pi i\mu_n})$, 
$\cdots\cdots$, $\cdots\cdots$, $P_{n-1}=\diag(e^{2\pi i\mu_1},\cdots,
e^{2\pi i\mu_{n-2}},e^{\pi i(\mu_{n-1}+\mu_n)},e^{\pi i(\mu_{n-1}+\mu_n)})$,
$P_n=\diag(-e^{\pi i(\mu_1+\mu_n)},e^{2\pi i\mu_2},\cdots,e^{2\pi i\mu_{n-1}},
-e^{\pi i(\mu_1+\mu_n)})$. Note that $P_n$ can also be written as
$P_n=\diag(e^{\pi i(\mu_n+\mu_{n+1})},e^{2\pi i\mu_2},\cdots,
e^{2\pi i\mu_{n-1}},e^{\pi i(\mu_n+\mu_{n+1})})$.

\section{Discussion}
As we have seen, an instanton has BPS monopoles as constituents, and the
explicit results in eq.~(19) also easily reveal the abelian limit. As long 
as\footnote{The Harrington-Shepard solution~\cite{HaSh} with trivial 
holonomy (all $\mu_i=0$), can be reinterpreted as a bound state of a massive 
and a massless constituent BPS monopole. Massless monopole constituents, 
giving rise to so-called non-abelian clouds, may play an important role in 
electric-magnetic duality as the dual of gluons~\cite{Wein}. These massless 
constituents are delocalised and have no obvious abelian limit.} all
$\nu_m\neq 0$, the field outside the core of both monopoles is indeed 
that of two ``BPS'' Dirac monopoles (i.e. dyons). For $SU(2)$ outside the 
cores, i.e. assuming $r_m\nu_m\gg1$ for all $m$, one has $\tilde\chi=0$ and 
\beq
\phi(x)=\frac{|\vec x-\vec y_1|+|\vec x-\vec y_2|+|\vec y_2-\vec y_1|}{
|\vec x-\vec y_1|+|\vec x-\vec y_2|-|\vec y_2-\vec y_1|}.
\eeq
We note that when {\em neglecting} the exponential corrections, $\phi^{-1}(x)$ 
vanishes on the line connecting the two constituents, and $\log\phi(x)$ is
harmonic outside of this line. The singularity represents the return flux 
of the Dirac monopole pair, described in the abelian limit by a term 
proportional to $\partial_j^2\log\phi(x)$ in the magnetic field
\beq
E_k=\frac{i}{2}\tau_3\partial_k\partial_3\log\phi,\quad
B_k=\frac{i}{2}\tau_3\left(\partial_k\partial_3
\log\phi-\delta_{k3}\partial_j^2\log\phi\right).
\eeq
In the full theory the return flux is absent (indeed $\phi^{-1}(x)$ has only 
an isolated zero at $x=0$, corresponding to a gauge singularity~\cite{KrvB}).

\begin{figure}[htb]
\vspace{3.3cm}
\includegraphics{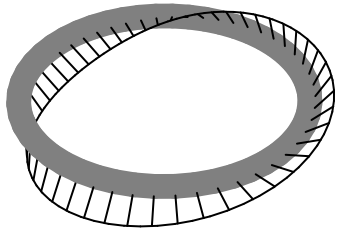}
\caption{A closed monopole line, rotating its frame when completing 
the circle. The topological charge is given by the net number of windings
(here one) of the frame.}
\end{figure}

As was already emphasised in ref.~\cite{LAT97}, to understand how in 
the abelian limit the topological charge can be recovered, one needs
to preserve the Taubes-winding. This information is described by the 
framing of the core due to the charged components of the monopole field. 
Interestingly this describes a Hopf fibration~\cite{KrvB}, see figure 4.
Recently the role of the Taubes-winding and Hopf fibration for retrieving
topological charge was confirmed in great detail within the context of the
abelian projection~\cite{Jahn}.

We have also seen that the fermion zero-mode is localised on the constituent 
with the Taubes-winding and this makes it likely to conjecture that this 
holds for general monopole loops that support non-trivial Taubes-winding, 
which may be relevant to understanding chiral symmetry breaking in the 
context of monopole degrees of freedom. The question which field 
configurations are more important, is however a bit like the chicken and 
the egg problem. After all, we can now make instantons out of monopoles 
{\em and} monopoles out of instantons. Monopoles are typically used for 
describing the confining large distance behaviour. Recently it has, 
however, been pointed out there can be a hidden large size instanton 
component hitherto undetected~\cite{Dist}. Much remains to be done here, 
but the calorons have provided us interesting new avenues to follow.

\begin{figure}[htb]
\vspace{5.5cm}
\includegraphics{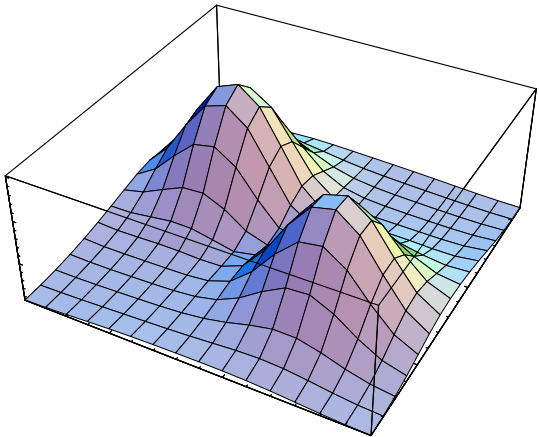}
\caption{Lattice caloron profile created with improved cooling on a 
$16^3\times 4$ lattice with twist in the time direction. The total action 
is $1.000185\times8\pi^2$. Vertically is plotted $\log(1+s/3)$, with $s$ 
the action density at the lattice site (after clover averaging), through 
a plane near to both constituent locations~[2].}
\end{figure}

Let us end by mentioning that the calorons with non-trivial holonomy have 
also been found on a finite lattice. For $SU(2)$ with twisted-boundary 
conditions remarkably it was found that when the twist is in the time 
direction, the constituent locations are free, but both constituents have 
equal mass, see figure 5. Whereas, when the twist is in the space direction, 
the constituent locations are maximally separated in the direction of the 
magnetic flux, but the mass ratios can be arbitrary~\cite{MTAP}. This has 
been understood in the general context of the Nahm transformation for the 
torus, relating finite volume calorons to finite volume 
instantons~\cite{Tdual}. The suggestion of ref.~\cite{KrvB} to find these 
calorons on the lattice by freezing the links on the boundary of the lattice 
to enforce the proper holonomy has also been realised recently~\cite{Ves}.

\section*{Acknowledgements}
I am grateful to the organisers, and in particular to Valya Mitrjushkin, for 
their invitation and for the wonderful atmosphere in Dubna. I thank all my 
collaborators of the work presented here, for having contributed to the 
acquired insights. Discussions with participants, in particular with Maxim 
Chernodub, Michael M\"uller-Preussker and Sasha Veselov are greatly 
appreciated. This work was supported in part by a grant from ``Stichting 
Nationale Computer Faciliteiten (NCF)'' for use of the Cray Y-MP C90 at SARA.

\end{document}